\documentstyle[12pt]{article}
\begin{document}
\baselineskip=15pt

\newcommand{\be}{\begin{equation}}
\newcommand{\ee}{\end{equation}}
\newcommand{\bq}{\begin{eqnarray}}
\newcommand{\eq}{\end{eqnarray}}
\newcommand{\x}{{\bf x}}
\newcommand{\p}{\varphi}
\newcommand{\Sc}{Schr\"odinger\,}
\newcommand{\del}{\nabla}
\newcommand{\zm}{z^{\prime\mu}}
\newcommand{\zn}{z^{\prime\nu}}
\newcommand{\zl}{z^{\prime\lambda}}
\newcommand{\zr}{z^{\prime\rho}}
\newcommand{\Sp}{{\,\,\,\,\,\,\,\,\,\,\,\,\,}}

\newcommand{\s}{\mbox{O(N) $\sigma$-model\,}}

\begin{titlepage}
\rightline{DTP 95/45}
\vskip1in
\begin{center}
{\LARGE The \s Laplacian}
\end{center}
\vskip1in
\begin{center}
{\large
Paul Mansfield

and Jiannis Pachos

Department of Mathematical Sciences

University of Durham

South Road

Durham, DH1 3LE, England}

{\it P.R.W.Mansfield@durham.ac.uk}

{\it Jiannis.Pachos@durham.ac.uk}
\end{center}
\vskip1in
\begin{abstract}
\noindent
For fields that vary slowly on the scale of the lightest mass the
logarithm of the vacuum functional of a massive quantum field theory
can be expanded in terms of local functionals satisfying a form of
the \Sc equation, the principal ingredient of which is a regulated
functional Laplacian.
We construct to leading order a Laplacian for the \s that acts on
such local functionals. It is determined by imposing rotational
invariance in the internal space together with closure of the
Poincar\'e algebra.

 \end{abstract}

\end{titlepage}


\section{\bf  The \s Laplacian}

The \s shares many features with Yang-Mills theory. They are both
classically conformally invariant but generate mass quantum
mechanically. Both are renormalisable \cite{bre}, asymptotically
free, and have large-N expansions
\cite{Polyakov}. However the \s, has the advantage of tractability
so that,
for example, mass generation can be explicitly demonstrated within
the large-N
expansion. This makes it a useful laboratory to test techniques that
are ultimately intended for the study of Yang-Mills theory.
Recently one of us  proposed a new approach to the eigenvalue
problem for the Hamiltonians of massive quantum field theories
that should be applicable to theories that are classically
 massless \cite{Paul}. This
is based on a version of the  functional \Sc equation that acts
 directly on a local expansion
of the vacuum functional. (The \Sc representation approach to
field theory is discussed in \cite{Sym}-\cite {Horiguchi}).
This local expansion is applicable when the vacuum functional
is evaluated for fields that vary slowly on the scale of the inverse
of the mass. The purpose of this paper is to construct to
leading order the
functional Laplacian for the \s that acts on local functionals,
which is the principal ingredient in this approach. We will show
that this operator is determined by the symmetries of the model,
namely Poincar\'e invariance and
rotational symmetry in the internal space.

\bigskip
Consider the quantum mechanics of a non-relativistic particle of
mass, $m$, moving on the N-dimensional sphere with co-ordinates
$z^\mu(\tau)$ at time $\tau$.
The action is $S={m\over 2}\int d\tau g_{\mu\nu}{\dot z}^\mu
{\dot z}^\nu$ where
$g_{\mu\nu}$ is the metric on the sphere which we take to have
radius $a$.
The Hamiltonian in the \Sc representation is $H=-{1\over 2m}\Delta$,
where $\Delta$ is the Laplacian, and the eigenfunctions are the
spherical harmonics.
The \s can be thought of as a
natural generalisation in which we replace the particle by a
curve parametrised by $\sigma$, say, and choose an action that
is relativistically invariant
in the space-time $(\sigma,\tau)$ so that
\be
S={1\over 2\alpha}\int d\sigma d\tau\, g_{\mu\nu}\left({\dot z}^\mu
{\dot z}^\nu
-{ z}^{\prime\mu}{ z}^{\prime\nu}\right)
\ee
where $\prime$ and $\cdot$ denote differentiation with respect to
$\sigma$ and $\tau$, and $\alpha$ is a coupling constant.
Formally, the Hamiltonian in the \Sc representation would be
\be
H=-{\alpha\over 2}\Delta+{1\over 2\alpha}\int d\sigma g_{\mu\nu}\,
{ z}^{\prime\mu}{z}^{\prime\nu}
\label{eq:ham}
\ee
where $\Delta$ is now the Laplacian constructed from an inner
product on variations of the co-ordinates
\be
(\delta z,\delta z)=\int d\sigma\, g_{\mu\nu}\,{\delta z}^\mu
{\delta z}^\nu
\ee
This is a metric with an infinite number of components labelled
by
$\sigma$ and also $\mu,\nu$. If we use bold-face type to denote
infinite
component tensors and ordinary type to denote tensors on $S^N$
then
${\bf g}_{\mu_1\mu_2}(\sigma_1,\sigma_2)=g_{\mu_1\mu_2}
(z(\sigma_1))\,\delta (\sigma_1-
\sigma_2)$. The pair $(\mu_1,\sigma_1)$ should be treated
as a single index, as should $(\mu_2,\sigma_2)$, so that
$\bf g$ is a two-index tensor, as is usual.
This has an inverse, ${\bf g}^{\mu_1\mu_2}(\sigma_1,\sigma_2)=
g^{\mu_1\mu_2}(z(\sigma_1))\,\delta (\sigma_1-
\sigma_2)$. The infinite dimensional Kronecker delta
${\bf I}^{\mu_1}_{\mu_2}(\sigma_1,\sigma_2)=
\delta ^{\mu_1}_{\mu_2}\delta(\sigma_1-\sigma_2)$,
and is equal to the functional derivative of
$z^{\mu_1}(\sigma_1)$
with respect to $z^{\mu_2}(\sigma_2)$. It is tempting
to try to generalise Riemannian geometry to this infinite
dimensional case, so that ${\bf g}$ and ${\bf I}$ transform
covariantly under general co-ordinate transformations in
which $z^{\mu}(\sigma)\rightarrow {\tilde z}^{\mu}(\sigma)$
and $\tilde z$ is a functional of $z$ i.e. it depends
on the entire curve $z=z(\sigma)$.
However, the utility of this for quantum field theory is
not clear. What is important is that the theory be
invariant under rotations in the internal space, that is
to say the isometries of the finite dimensional metric
$g_{\mu\nu}$, since these underpin the renormalisability
of the theory \cite{bre}. These are rigid co-ordinate
transformations, so we will restrict our attention to
co-ordinate transformations in which $\tilde z^\mu (\sigma)$
is a function of the co-ordinates $z^\nu (\sigma)$ i.e.
just at the point $\sigma$ on the curve.
Under this restricted class of transformations a
finite-dimensional vector
$V^\mu (z)$ on $S^N$, e.g. $\zm$,
 may be thought of as an infinite dimensional vector
${\bf V}^\mu (\sigma)=V^\mu (z(\sigma))$.
To construct the Laplacian we need a covariant derivative.
Given the infinite dimensional metric we can follow the
usual construction of the Levi-Civita connection, ${\bf D}$,
which will transform covariantly under general co-ordinate
transformations and therefore under our restricted
transformations. Thus
\be
{\bf D}_{\mu_2}(\sigma_2){\bf V}^{\mu_1}(\sigma_1)
={\delta{\bf V}^{\mu_1}(\sigma_1)\over\delta z^{\mu_2}
(\sigma_2)}+
\int d\sigma_3\,{\bf \Gamma}^{\mu_1}_{\mu_2 \mu_3}
(\sigma_1,\sigma_2,\sigma_3)
{\bf V}^{\mu_3}(\sigma_3)
\ee
where the infinite dimensional Christoffel symbol is
related to that on $S^N$ by \be
{\bf \Gamma}^{\mu_1}_{\mu_2\mu_3}(\sigma_1,\sigma_2,\sigma_3)=
\delta(\sigma_1-\sigma_2)\,\delta(\sigma_2-\sigma_3)\,
{ \Gamma}^{\mu_1}_{\mu_2\mu_3}(z(\sigma_1))
\ee
If we apply this to a vector that depends on $\sigma$
and $z(\sigma)$ but not on its derivatives (we will
call this property ultra-locality) then it is
straightforward to compute
\be
{\bf D}_{\mu_2}(\sigma_2){\bf V}^{\mu_1}(\sigma_1)
=(D_{\mu_2}V^{\mu_1})|_{z(\sigma_1)}\,\delta(\sigma_1
-\sigma_2),
\ee
where $D$ is the covariant derivative on $S^N$.
Similarly we can easily compute the covariant derivative
of ${\bf z}^{\prime\mu}$ as
\be
{\bf D}_{\mu_2}(\sigma_2)\,{\bf z^\prime}^{\mu_1}(\sigma_1)=
\delta^{\mu_1}_{\mu_2}\,\delta^\prime(\sigma_1-\sigma_2)+
\left(\Gamma ^{\mu_1}_{\mu_2 \rho}z^{\prime\rho}
\right)|_{z(\sigma_1)}
\,\delta(\sigma_1-\sigma_2)
\ee
with $\Gamma$ the finite dimensional Christoffel symbol.
Now the finite dimensional intrinsic derivative ${\cal D}
={\partial\over\partial\sigma}
+\zm D_\mu$ maps finite dimensional vectors to finite
dimensional vectors,
so we can use it to define new infinite component vectors as
${\cal D}|_\sigma {\bf V}^\mu (\sigma)\equiv({\cal D}V)^\mu
|_{z(\sigma)}$.
The infinite dimensional tensor $ {\bf I}_{\mu_2}^{\mu_1}
(\sigma_1,\sigma_2)$,
thought of in terms of finite dimensional tensors,
is an element of the product of the tangent space at
$z(\sigma_1)$
and the co-tangent space at $z(\sigma_2)$. If we apply
the intrinsic derivative
with respect to $\sigma_1$ then it acts only on the
$\mu_1$ index to give
\be
 {\cal D}|_{\sigma_1 }{\bf I}_{\mu_2}^{\mu_1}(\sigma_1,
\sigma_2)
 =\delta^{\mu_1}_{\mu_2}\,\delta^\prime(\sigma_1-\sigma_2)+
\left(\Gamma ^{\mu_1}_{\mu_2 \rho}z^{\prime\rho}\right)
|_{z(\sigma_2)}
\,\delta(\sigma_1-\sigma_2)=
{\bf D}_{\mu_2}(\sigma_2)\,{\bf z^\prime}^{\mu_1}(\sigma_1).
\ee
Similarly the intrinsic derivative
with respect to $\sigma_2$ acts only on the $\mu_2$ index to
give
\be
 {\cal D}|_{\sigma_2 }{\bf I}_{\mu_2}^{\mu_1}(\sigma_1,
\sigma_2)=
-{\bf D}_{\mu_2}(\sigma_2)\,{\bf z^\prime}^{\mu_1}(\sigma_1)
\ee
Since $\zm ={\cal D}z^\mu (\sigma)$ this implies that
$[{\bf D},{\cal D}]z^\mu=0$, so that this commutator
also annihilates any ultra-local scalar. It will prove
useful to know the value of this commutator when it
acts on vectors. If $V^\mu$ is ultra-local then
\be
[{\bf D}_{\mu_1}(\sigma_1),\,{\cal D}|_{\sigma_2}]\,
{\bf V}^{\mu_2}(\sigma_2)
=z^{\prime\rho}(\sigma_1)\left( R_{\rho\mu_1 \lambda}
^{\Sp
\mu_2}V^\lambda
\right)|_{z(\sigma_1)}\,\delta(\sigma_1-\sigma_2)
\ee
where $R$ is the finite dimensional Riemann tensor
given by $[D_\mu, D_\lambda]\,
V^\beta= R_{\lambda \mu\rho}^{\Sp
\beta}V^\rho$, so that on the sphere  $R_{\alpha\beta
\gamma\delta}=(g_{\alpha\gamma}g_{\beta\delta}-g_{\beta\gamma}
g_{\alpha\delta})/a^2$.
(This computation is essentially the same as for the
finite dimensional case, decorated by delta-functions
so we omit the details).
For $\zm$ we obtain
\be
[{\bf D}_{\mu_1}(\sigma_1),{\cal D}|_{\sigma_2}]\,
{\bf z^\prime}^{\mu_2}(\sigma_2)
=z^{\prime\rho}(\sigma_1)z^{\prime\lambda}(\sigma_1)\,
R_{\mu_1 \rho\lambda}^{\Sp
\mu_2}|_{z(\sigma_1)}\,\delta(\sigma_1-\sigma_2).
\ee

\bigskip
The obvious definition of the Laplacian as the second
covariant derivative
with indices contracted using ${\bf g}$
does not exist because the
two functional derivatives act at the same value of
$\sigma$
and also because the determinant of the infinite
dimensional metric $\bf g$ is ill-defined. Instead we
will look for a regulated expression of the form
\be
{\Delta}_s=\int d\sigma_1 d\sigma_2\,{\bf G}^{\mu_1\mu_2}
(\sigma_1,\sigma_2
;s)\,
{\bf D}_{\mu_1}(\sigma_1){\bf D}_{\mu_2}(\sigma_2).
\ee
The kernel, ${\bf G}$, is constrained by a number of
physical requirements.
We will see that these are sufficent to determine its
form, at least to leading order, when the Laplacian
acts on local functionals.
Firstly we require that it is a regularisation of the
inverse metric, so we
will assume that it depends on a cut-off parameter,
$s$, with the dimensions of squared length and takes
the form$
{\bf G}^{\mu_1\mu_2}(\sigma_1,\sigma_2;s)={\cal G}_s
(\sigma_1-\sigma_2)\,
{ K}^{\mu_1\mu_2}(\sigma_1,\sigma_2)$,
where ${\cal G}_s(\sigma_1-\sigma_2)\sim \delta
(\sigma_1-\sigma_2)$
as $s\downarrow 0$, and ${ K}$ is a non-singular
functional of $z^\mu$
included so that  ${\bf G}$ transforms as a tensor
under the restricted class of co-ordinate transformations.
Our problem is to construct $K$.
For $\bf G$ to be a regularisation of $\bf g$
we require that ${ K}^{\mu_1\mu_2}(\sigma,\sigma)=
g^{\mu_1\mu_2}(z(\sigma))$.
Finally, since we work in a Hamiltonian formalism,
Poincar\'e invariance is not manifest and must be
imposed by demanding that the
generators of these transformations satisfy the
Poincar\'e algebra. This
is the integrability condition for the action of
the generators on wave-functionals, $\Psi$.
Ignoring regularisation the Poincar\'e  generators
are the Hamiltonian,
(\ref{eq:ham}), the momentum $P=\int d\sigma\zm
{\bf D}_\lambda(\sigma)$ and
the Lorentz generator $L=-\alpha M+\alpha^{-1}N$,
where
\be
M={1\over 2}\int d\sigma\,\sigma g^{\mu_1\mu_2}\,
{\bf D}_{\mu_1}(\sigma){\bf D}_{\mu_2}(\sigma),\quad
N={1\over 2}\int d\sigma\,\sigma
g_{\mu_1\mu_2}z^{\prime \mu_1}z^{\prime\mu_2}
\ee
Again ignoring problems of regularisation, these
satisfy the Poincar\'e
algebra
\be
[P,H]=0,\quad [L,P]=H,\quad [L,H]=P.
\label{eq:cond}
\ee
We require that this algebra still holds when
regulators are in place.
The momentum operator does not need to be regulated.
We regulate the Laplacian as above to yield a cut-off
Hamiltonian $H_s$.
In scalar $\varphi^4$ theory it was shown by Symanzik
\cite{Sym} that in the
\Sc representation
wave-functionals expressed in terms of renormalised
fields
have a finite limit as the regulator is removed, and
since the Hamiltonian generates displacements in
$\tau$ it has a finite action on these wave-functionals.
(The field undergoes an additional renormalisation for
$\varphi^4$ due to boundary effects but these are
presumably absent for the $O(N)$ sigma-model due
to rotational invariance in the internal space).
Thus the
limit as $s\downarrow 0$ of $H_s\Psi$ exists and
is what we mean by the Hamiltonian applied to $\Psi$.
We assume that this carries over to the
O(N) sigma-model. Similarly we introduce a cut-off
into $L$ to obtain $L_s$ which should have a finite
limit when applied to wave-functionals.
The commutator $[L,P]=H$
implies that $L$ should be regulated with the same
kernel as $H$, so we replace the operator $M$ by
\be
\int d\sigma_1d\sigma_2\,{\sigma_1+\sigma_2\over 2}
\,{\bf G}_s^{\mu_1\mu_2}(\sigma_1,\sigma_2)\,
{\bf D}_{\mu_1}(\sigma_1){\bf D}_{\mu_2}(\sigma_2)
\equiv M_s
\ee

\bigskip
    Ultimately we are interested in constructing
the \Sc equation for an expansion of the vacuum-functional
in terms of local functionals, i.e. integrals of functions
of $z(\sigma)$ and a finite number of its derivatives at
the point
$\sigma$.
So we consider the conditions on the kernel $\bf G$ that
arise from applying
(\ref{eq:cond}) to such test functionals.
It is convenient to order such test functionals according
to the powers of $\cal D$ for the following reason. If we
act with $\Delta_s$ on a functional of the form
$\int d\sigma\,f(z(\sigma),\sigma)_{\mu_1..\mu_n}
z^{\prime\mu_1}..z^{\prime\mu_n}\equiv F_n$, where $f$ is
ultra-local, then the two functional derivatives in the
Laplacian will act on the $z^{\prime\mu_1}..z^{\prime\mu_n}$
to generate a second order differential operator acting on
$\delta(\sigma_1-\sigma_2)$.
Integrating by parts allows this operator to act on one of
the $\sigma$ arguments of the kernel, whilst the
delta-function sets both arguments equal.
The consequence of this is that ${\Delta_{s}}F_n$
depends on the second derivative of the kernel evaluated
at co-incident points.
Demanding the  closure of the Lorentz algebra acting
on $F_n$
will constrain this quantity, whereas if we consider the
Laplacian
applied
to a test-functional containing higher derivatives we
obtain information about the higher derivatives of the
kernel. $F_n$ is the lowest order functional that gives
a constraint. We treat $F_n$ as a scalar so that
\be
{\bf D}_{\mu}(\sigma)F_n={\delta F_n\over \delta z^\mu
(\sigma)}=
\left(D_\mu \,f_{\mu_1..\mu_n}
z^{\prime\mu_1}..z^{\prime\mu_n}-n{\cal D}(f_{\mu\mu_2..
\mu_n}
z^{\prime\mu_2}..z^{\prime\mu_n})\right)|_\sigma
\ee
which is an infinite component co-vector. Using the
commutators of ${\bf D}$
and ${\cal D}$ worked out above it is easy to show that
\bq
\Delta_s F_n=&&\int d\sigma\, {\bf G}^{\mu\nu}
(\sigma,\sigma)\left(D_\mu D_\nu f_{\rho_1..\rho_n}
-nR_{\mu\rho_1\nu}^{\Sp \lambda}\,f_{\lambda\rho_2..
\rho_n}\right)z^{\prime\rho_1}..z^{\prime \rho_n}
\nonumber\\
&&
+n\int d\sigma\,   \left( ({\cal D}|_\sigma +
{\cal D}|_{\sigma^\prime})
{\bf G}^{\mu\nu}(\sigma,\sigma^\prime) \right)
|_{\sigma=\sigma^\prime}
D_\mu  f_{\nu\rho_2..\rho_n}z^{\prime\rho_2}..
z^{\prime \rho_n}
\nonumber\\
&&+n(n-1)\int d\sigma\left({\cal D}|_{\sigma}
{\cal D}|_{\sigma^\prime}\,{\bf G}^{\mu\nu}
(\sigma,\sigma^\prime)\right)_{\sigma=\sigma^\prime}
f_{\mu\nu\rho_3..\rho_n}z^{\prime\rho_3}..
z^{\prime \rho_n}
\label{eq:delF}
\eq
Given the dimension of ${\bf G}$, (inverse length),
and its transformation properties we can set
\be
{\bf G}^{\mu\nu}(\sigma,\sigma)={1\over \sqrt s}
\left(\,b_0^0\,g^{\mu\nu}\right)
\ee
so that
\be
\left( ({\cal D}|_\sigma +{\cal D}|_{\sigma^\prime})
{\bf G}^{\mu\nu}(\sigma,\sigma^\prime)\right)_
{\sigma=\sigma^\prime}
={\cal D}\left(   {1\over \sqrt s}\left(\,b_0^0\,
g^{\mu\nu} \right)\right)= 0
\ee
and
\be
\left(
{\cal D}|_{\sigma}{\cal D}|_{\sigma^\prime}\,
{\bf G}^{\mu\nu}(\sigma,\sigma^\prime)\right)_{\sigma=
\sigma^\prime}
=-{1\over {\sqrt s}^3}\left(\,b_0^1\,g^{\mu\nu}+
sb_1^1 \,g_{\lambda\rho}z^{\prime\lambda}z^{\prime\rho}
\,g^{\mu\nu}
+sb^1_2\,z^{\prime\mu}z^{\prime\nu}
\right)
\ee
where $b^0_1,b_0^1,b_1^1,b_2^1..$ are dimensionless
constants.
$b_0^0$ and $b_0^1$ are determined by our choice of
regularisation of the delta-function, ${\cal G}_s$
\be
b_0^0=\sqrt s\, {\cal G}_s(0),\quad b_0^1={\sqrt s}^3
\,{\cal G}^{\prime\prime}_s(0),
\ee
our problem is to relate them to the remaining
coefficents by imposing the closure of the Poincar\'e
algebra.  Using these expressions we can write
$\Delta_s F_n$ as
\be
\Delta_s F_n=-{n(n-1)\,b_0^1\over{\sqrt s}^3}\int
d\sigma\,g^{\mu\nu}f_{\mu\nu\rho_3..\rho_n}\,
z^{\prime\rho_3}..z^{\prime \rho_n}
+{1\over\sqrt s}\int d\sigma
\left( J_nf\right)_{\rho_1..\rho_n}\,
z^{\prime\rho_1}..z^{\prime \rho_n}
\label{eq:aaaa}
\ee
where
\bq
\left( J_nf\right)_{\rho_1..\rho_n}
&=&
b^0_0 g^{\mu\nu}\left( D_\mu D_\nu\, f_{\rho_1..\rho_n}
+n\,f_{\lambda(\rho_2..\rho_n}R_{\rho_1)\mu\nu}^{\Sp
\lambda}\right)
\nonumber\\
&&
-n(n-1)\left(b_1^1g^{\mu\nu}\,f_{\mu\nu (\rho_3..\rho_n}
g_{\rho_1\rho_2)}
+b_2^1\,f_{\rho_1..\rho_n}\right),\label{eq:J}
\eq
and bracketed indices are symmetrised.
The calculation of $M_sF_n$ is essentially the same,
but with ${\sigma_1+\sigma_2\over 2}{\bf G}$ replacing
${\bf G}$,  so that there is an additional piece coming
from the second integral on the right-hand side of
(\ref{eq:delF})
\bq
M_s F_n&=&-{n(n-1)\,b_0^1\over{\sqrt s}^3}\int d\sigma\,
\sigma\,g^{\mu\nu}f_{\mu\nu\rho_3..\rho_n}\,
z^{\prime\rho_3}..z^{\prime \rho_n}\nonumber\\
&&+{1\over\sqrt s}\int d\sigma\,\sigma\, \left( J_nf
\right)_{\rho_1..\rho_n}\,
z^{\prime\rho_1}..z^{\prime \rho_n}
+{n\,b_0^0\over\sqrt s}\int d\sigma \,D^\mu f_{\mu\rho_2..
\rho_n}z^{\prime\rho_2}..z^{\prime \rho_n}
\label{eq:bbbb}
\eq

\bigskip
In \cite{Paul} it was shown  that for scalar field theory
the expansion of the vacuum functional in terms of local
functionals, which is valid for slowly varying fields,
does not satisfy the
obvious \Sc equation because expanding in terms of local
functionals does not commute with removing the cut-off.
However, by studying the analyticity properties of the
Laplacian it was shown how to re-sum the cut-off dependence
of $\Delta_s\Psi$ so
as to be able to remove the cut-off correctly. This
re-summation
is accomplished by performing a contour integral over
$s$ and has the effect of replacing
$s^{-n}$ in $\Delta_s\Psi$ by $\lambda^n c(n)$ where
$c(n)$ is just a numerical function of $n$ and $\lambda$
is a new cut-off that is to be taken to infinity.
We shall assume that such a re-summation may be performed
here.

\bigskip
      We will now consider the conditions placed on the
kernel by demanding that the Poincar\'e algebra close
when acting on the local functionals that we hope to
construct the wave functionals from. We require that
$([L,H]-P)F_n=0$.
Introducing regulators into the generators we require
that as $s_1,s_2\downarrow 0$
\be
{\textstyle 1\over 4}[- \alpha M_{s_1}+\alpha^{-1}N,\,
-\alpha\Delta_{s_2}+\alpha^{-1}V]F_n =PF_n
\label{eq:alg}
\ee
Contributing to this equation there will be a number
of terms in which the
removal of the cut-off simply replaces the kernel $\bf G$
by the metric $\bf g$ without any singularity arising.
By themselves the sum of such terms satisfies the
equation since they are just what would occur if we
ignored the problem of regularisation altogether.
The remaining terms have to cancel against each
other to satisfy
(\ref{eq:alg}). These are the terms that, in the
absence of a regulator, involve two functional
derivatives at the same point acting on a single
local functional. Thus we require $[M_{s_1},
\Delta_{s_2}]F_n=0$ as well as
$M_{s_1}V=0$ and $\Delta_{s_2}N=0$.
Using the results above we compute
\bq
&&[M_{s_1},\Delta_{s_2}]F_n=\nonumber\\
&&(s_2^{-1/2}s_1^{-3/2}-   s_2^{-3/2}s_1^{-1/2} )
\,2n(n-1)\,b_0^1
\left(kb^0_0+(N-1)b^1_1+(4n-6)(b^1_1+b^1_2)
\right)\times
\nonumber\\
&&\quad\quad\int d\sigma\,\sigma\, (tr\, f)_{\rho_3
..\rho_n}
z^{\prime\rho_3}..z^{\prime \rho_n}\nonumber\\
&&-2n(n-1)\,b^0_0(s_1s_2)^{-1/2}\int d\sigma\Biggl(
\left(b_1^1+      {kb^0_0\over N-1}\right)D_{(\rho_2}
(tr\,f)_{\rho_3..\rho_n)}
\nonumber\\
&&
\quad\quad+\left(b_2^1-{kb^0_0\over N-1}\right)D^\mu
f_{\mu\rho_2..\rho_n}
\Biggr)z^{\prime\rho_2}..z^{\prime \rho_n}
\eq
where $
(tr\,f)_{\rho_3..\rho_n}=g^{\mu\nu}f_{\mu\nu\rho_3..
\rho_n}$.
This will vanish for all
$n$ by taking
\be
b_1^1=-b_2^1=-{kb^0_0\over N-1}=-{b^0_0\over a^2}
\label{eq:conditions}
\ee
Finally, consider the conditions $M_{s}V=0$ and
$\Delta_{s}N=0$.
Using (\ref{eq:aaaa}) and (\ref{eq:bbbb}) we obtain
\be
M_s\,V=\Delta_s\,N=-{2Nb_0^1\over{\sqrt s}^3}\int
d\sigma\,\sigma
-{2\over\sqrt s}(kb_0^0+Nb_1^1+b^1_2)\int d\sigma\,
\sigma\,g_{\mu\nu}
\,z^{\prime\mu}z^{\prime\nu},
\ee
so that these conditions are also satisfied by
(\ref{eq:conditions})
if we take the ill-defined integral $\int d\sigma\,
\sigma$ to vanish
on the grounds that the integrand is odd. (We note
in passing that $\Delta_s V$
is just a constant.)

\bigskip
   Substituting our results back into $\bf G$ we obtain
the following information
\be
{\bf G}^{\mu\nu}(\sigma,\sigma)={\cal G}_s(0)\,g^{\mu\nu}
\ee
and
\be
\left({\cal D}|_\sigma{\cal D}|_{\sigma^\prime}
{\bf G}^{\mu\nu}(\sigma,\sigma^\prime)\right)|_{\sigma
=\sigma^\prime}
=-{\cal G}^{\prime\prime}_s(0)\,g^{\mu\nu}
+{\cal G}_s(0)\,R^{\,\mu\,\,\,\nu}_{\,\,\,\,\lambda
\,\,\,\rho}z^{\prime\lambda}
z^{\prime\rho}
\label{eq:last}
\ee
To these we can add the condition $\left({\cal D}
|_\sigma{\bf G}^{\mu\nu}(\sigma,\sigma^\prime)\right)
|_{\sigma=\sigma^\prime}=0$, which follows from dimensional
analysis and rotational invariance.
These results can be used to re-construct $\bf G$ at
non-coincident points
in a Taylor expansion. To do this covariantly we
introduce ${\bf W}^{\mu_1}
_{\mu_2}(\sigma_1,\sigma_2)$ defined by
${\cal D}|_{\sigma_1}{\bf W}^{\mu_1}
_{\mu_2}(\sigma_1,\sigma_2)=0,$ and ${\bf W}^{\mu_1}
_{\mu_2}(\sigma,\sigma)=\delta^{\mu_1}_{\mu_2}$,
so that $\bf W$ is the path-ordered exponential
integral of the Christoffel
symbols. It is invertible as an $N\times N$ matrix,
so we can set
${\bf G}^{\mu_1\mu_2}(\sigma_1,\sigma_2)={\bf W}^{\mu_1}
_{\nu}(\sigma_1,\sigma_2)\,T^{\nu\mu_2}(\sigma_1,
\sigma_2)$ where $T$ is a finite dimensional tensor
at $z(\sigma_2)$ but depends on $\sigma_1$. Now
$({\cal D}|_{\sigma_1})^n{\bf G}^{\mu_1\mu_2}
(\sigma_1,\sigma_2)=
{\bf W}^{\mu_1}
_{\nu}(\sigma_1,\sigma_2)({\partial\over\partial
\sigma_1})^n
T^{\nu\mu_2}(\sigma_1,\sigma_2)$, so that when we
set $\sigma_1$ and $\sigma_2$ equal the intrinsic
derivatives of $\bf G$ reduce to the ordinary
derivatives of $T$, enabling
us to use the usual Taylor expansion to find the
$\sigma_1$-dependence of
$T$.

\bigskip
      In conclusion, we have sought a Laplacian for
the \s of the form
\be
{\Delta}_s=\int d\sigma_1 d\sigma_2\,
{\bf G}^{\mu_1\mu_2}(\sigma_1,\sigma_2)\,
{\bf D}_{\mu_1}(\sigma_1){\bf D}_{\mu_2}(\sigma_2).
\ee
where the kernel $\bf G$ is a regularisation of
the infinite-dimensional
metric ${\bf g}^{\mu_1\mu_2}(\sigma_1,\sigma_2)=
{ g}^{\mu_1\mu_2}\delta(\sigma_1-\sigma_2)$. It
takes the form ${\cal G}_s(\sigma_1-\sigma_2)\,
K^{\mu_1\mu_2}(\sigma_1,\sigma_2)$ where ${\cal G}$
is a regularisation of a delta-function. By demanding
manifest invariance of $\Delta_s$ under rotations
in internal space
and the closure of the Poincar\'e algebra acting on
local functionals of the
form $\int d\sigma\,f(z(\sigma),\sigma)_{\mu_1..\mu_n}
z^{\prime\mu_1}..z^{\prime\mu_n}\equiv F_n$ we
obtained a number of conditions on $\bf G$ that were
all satisfied by (\ref{eq:conditions}). These
translate into the statement that up to terms of
$O\left((\sigma_1-\sigma_2)^3\right)$
\be
K^{\mu_1\mu_2}(\sigma_1,\sigma_2)={\bf W}^{\mu_1}_\nu
(\sigma_1,\sigma_2)
\left(g^{\nu\mu_2}|_{z(\sigma_2)}-{\textstyle
{1\over 2}}{(\sigma_1-\sigma_2)^2}
R^{\nu\,\,\,\mu_2}_{\,\,\,\lambda\,\,\,\rho}
z^{\prime\lambda}
z^{\prime\rho}|_{z(\sigma_2)}\right)
\ee
Given a choice of ${\cal G}_s$ these conditions
are sufficent to fix the action of $\Delta_s$ on
$F_n$. By considering the closure of the Poincar\'e
algebra acting on test functionals containing higher
derivatives of the co-ordinates we would obtain
information about the higher terms in the Taylor
expansion of $K$.

\bigskip
\noindent
Finally, PM would like to acknowledge a grant
from the
Nuffield Foundation, and JP a studentship from
the University of Durham.



\end{document}